\documentclass[]{aastex}

\newcommand{\myemail}{cwalsh@strw.leidenuniv.nl}

\usepackage{amsmath}
\usepackage{amssymb}
\usepackage{url}
\usepackage{gensymb}
\usepackage{graphics}
\usepackage{subfigure}
\usepackage{wasysym}
\usepackage{textcomp}
\usepackage[version=3]{mhchem}

\shorttitle{ALMA Observations of HD~100546}
\shortauthors{Catherine Walsh et al.}

\begin{document}

\title{ALMA HINTS AT THE PRESENCE OF TWO COMPANIONS IN THE DISK AROUND HD~100546}
\author{Catherine Walsh\altaffilmark{1}, 
Attila Juh\'{a}sz\altaffilmark{1}, 
Paola Pinilla\altaffilmark{1}, 
Daniel Harsono\altaffilmark{1,2}
Geoffrey~S.~Mathews\altaffilmark{1,3},
William~R.~F.~Dent\altaffilmark{4,5}  
Michiel~R.~Hogerheijde\altaffilmark{1},
T.~Birnstiel\altaffilmark{6},
Gwendolyn Meeus\altaffilmark{7}, 
Hideko Nomura\altaffilmark{8}, 
Yuri Aikawa\altaffilmark{9}, 
T.~J.~Millar\altaffilmark{10}, 
G\"{o}ran Sandell\altaffilmark{11}}

\altaffiltext{1}{Leiden Observatory, Leiden University, P.~O.~Box 9531, 2300~RA Leiden, The Netherlands}
\altaffiltext{2}{SRON Netherlands Institute for Space Research, PO Box 800, 9700 AV Groningen, The Netherlands}
\altaffiltext{3}{Department of Physics and Astronomy, University of Hawaii, 2505 Correa Rd., Honolulu, HI 96822, USA}
\altaffiltext{4}{Joint ALMA Observatory (JAO), Alonso de C\'{o}rdova 3107, Vitacura, Santiago, Chile}
\altaffiltext{5}{European Southern Observatory (ESO), Alonso de C\'{o}rdova 3107, Vitacura, Santiago, Chile}
\altaffiltext{6}{Harvard-Smithsonian Center for Astrophysics, 60 Garden Street, Cambridge, MA 02138, USA}
\altaffiltext{7}{Dep. de F\'{i}sica Te\'{o}rica, Fac.~de Ciencias, UAM Campus Cantoplanco, E-28049 Madrid, Spain}
\altaffiltext{8}{Department of Earth and Planetary Science, Tokyo Institute of Technology, 2-12-1 Ookayama, Meguro-ku, 152-8551 Tokyo, Japan}
\altaffiltext{9}{Department of Earth and Planetary Sciences, Kobe University, 1-1 Rokkodai-cho, Nada, 657-8501 Kobe, Japan}
\altaffiltext{10}{Astrophysics Research Centre, School of Mathematics and Physics, Queen's University Belfast, University Road, Belfast, BT7~1NN, UK}
\altaffiltext{11}{SOFIA-USRA, NASA Ames Research Center, MS 232-12, Building N232, Rm.~146, P.~O.~Box 1, Moffett Field, CA 94035-0001, USA}

\email{\myemail}

\begin{abstract}
HD~100546 is a well-studied Herbig~Be star-disk system 
that likely hosts a close-in companion with compelling observational evidence 
for an embedded protoplanet at 68~AU.  
We present ALMA observations of the HD~100546 disk which resolve the 
gas and dust structure at (sub)mm wavelengths.  
The CO emission (at 345.795~GHz) originates from an extensive molecular disk (390$\pm$20~AU in radius) whereas    
the continuum emission is more compact (230$\pm$20~AU in radius) suggesting radial drift of the mm-sized grains.  
The CO emission is similar in extent to scattered light images  
indicating well-mixed gas and $\mu$m-sized grains in the disk atmosphere.   
Assuming azimuthal symmetry, a single-component power-law model cannot reproduce 
the continuum visibilities.  
The visibilities and images are better reproduced by a double-component model: a
compact ring with a width of 21~AU centered at 26~AU and an outer ring with a width 
of 75$\pm$3~AU centered at 190$\pm$3~AU.
The influence of a companion and protoplanet on the dust evolution is investigated.   
The companion at 10~AU facilitates the accumulation of mm-sized grains within a compact ring, 
$\approx$~20--30~AU, by $\approx$~10~Myr.  
The injection of a protoplanet at 1~Myr hastens the ring formation ($\approx$~1.2~Myr) and also triggers the development 
of an outer ring ($\approx$~100--200~AU).  
These observations provide additional evidence for the presence of a close-in companion and 
hint at dynamical clearing by a protoplanet in the outer disk.
\end{abstract}

\keywords{protoplanetary disks --- stars: formation --- stars: individual (HD~100546) --- stars: pre-main sequence --- submillimeter: planetary systems}


\section{INTRODUCTION}
\label{introduction}

Transition disks (TDs) are important for studying the advanced stages 
of protoplanetary disk evolution \citep[see, e.g.,][]{espaillat14}.  
TDs were originally identified as sources for which the 
spectral energy distribution (SED) demonstrated a lack of near-infrared excess 
despite the presence of strong mid- to far-infrared excess. 
This was attributed to a gap in the inner disk 
devoid of small grains \citep[e.g.,][]{strom89}.  
Theory suggests gaps in TDs are cleared by close-in companions, with other disk-dispersal mechanisms, e.g., 
grain growth or photoevaporation, happening in parallel \citep[see, e.g.,][]{armitage11,williams11,espaillat14}.
SEDs provide {\em indirect} evidence of gaps in TDs; however, long-baseline 
interferometry at (sub)mm wavelengths has revealed their ring-like morphology \citep[e.g.,][]{andrews11}.     
The Atacama Large Millimeter/Submillimeter Array (ALMA) 
has revealed extreme asymmetries in the dust emission in several systems, 
indicative of dust traps triggered by the interaction 
between the disk and a close-in companion \citep{casassus12,vandermarel13} 
or gravitational instabilities \citep{fukagawa13}. 
ALMA observations have also demonstrated that gaps can contain a significant 
reservoir of molecular gas \citep{bruderer14}.

We present ALMA Cycle 0 observations 
of the transition disk encompassing HD~100546 which reveal the 
spatially-resolved gas and dust structure at (sub)mm wavelengths.  
\citet{pineda14} have already published these data; however, 
we reach different conclusions based on more thorough data processing.

\section{HD~100546}
\label{hd100546}

HD~100546 is a 2.4$\mathrm{M}_\odot$ B9V Herbig Be star located at 
103$\pm$6~pc which has a complex circumstellar environment \citep[e.g.,][]{vandenancker98,grady01}.  
Coronographic imaging show that the small grains extend to large radii ($\approx$~500~AU) 
and reveal evidence of spiral arms and disk brightness asymmetries 
\citep[e.g.,][]{pantin00,augereau01,grady01,ardila07,boccaletti13}.
SED models of the dust emission suggest a gap 
within $\approx$~10--13~AU  
and the presence of an inner tenuous dust disk, $\lesssim$~0.7~AU 
\citep{bouwman03,benisty10,tatulli11,panic14}. 
Observations of [OI] (6300~\AA) line emission and
OH and CO rovibrational transitions confirm the presence of residual 
gas close to the star with the observed dynamical perturbation of 
the gas likely induced by a massive close-in companion 
\citep{acke06,brittain09,vanderplas09,goto12,liskowsky12,brittain13,bertelsen14}.

Emission at 1.3 and 3.4~mm was detected using the 
Swedish-ESO 15~m Submillimeter Telescope (SEST) 
and the Australia Telescope Compact Array (ATCA)
yielding a total flux density of 465$\pm$20 and 36$\pm$3~mJy, respectively 
\citep{henning94,wilner03}. 
A plethora of molecular lines have been observed at far-infrared to 
(sub)mm wavelengths including emission from $^{12}$CO, $^{13}$CO, OH, and \ce{CH+} 
\citep[see, e.g.,][]{panic10,sturm10,thi11,meeus12,fedele13a}. 
These data have allowed contraints on the radial behaviour 
of the gas temperature structure, and indicate 
thermal decoupling of the gas and dust in the disk atmosphere 
\citep{bruderer12,fedele13b,meeus13}.

The detection of significant emission from a point source at a 
deprojected radius of 68$\pm$10~AU \citep{quanz13} is of utmost importance 
in indicating planet formation around HD~100546.
High-contrast angular differential imaging revealed that 
the source emission coincides with a reduction in surface 
brightness seen in corresponding polarimetric differential imaging \citep{quanz11}.  
\citet{quanz13} conclude that the most likely explanation is a young gas-giant planet 
(or protoplanet) caught in the act of formation, 
reasoning that a mature massive planet (coeval with the star) would have 
had sufficient time to significantly perturb the structure of the disk.  
    
\section{OBSERVATIONS}
\label{observations}

HD~100546 was observed during ALMA Cycle 0 operations on 2012 November 18 using 
24 antennas with baseline lengths between 21 and 375~m (program 2011.0.00863.S, P.~I. C.~Walsh).    
The source was observed in seven spectral windows in Band~7, each with a bandwidth 
of 469~MHz and a channel width of 0.122~MHz (0.24 and 0.21~km~s$^{-1}$ 
at 300 and 345~GHz, respectively, applying Hanning smoothing).  
The central frequencies in each spectral window are 
300.506, 301.286, and 303.927~GHz for the first execution, 
and 344.311, 345.798, 346.998, and 347.331~GHz for the second execution.  
The total on-source observation time was 13 and 14~mins, respectively.  
The data were calibrated using the Common Astronomy Software Package (CASA), version 3.4.
The quasar, 3C~297, was used as bandpass calibrator with Titan and a quasar, J1147-6753, 
used for amplitude and phase calibration, respectively.  
Self-calibration and imaging were performed using CASA version 4.1.  
During imaging it was noticed that the telescope pointing 
had not taken into account the proper motions of the source 
($\alpha_{2000}$~=~11$^\mathrm{h}$~33$^\mathrm{m}$~25\fs44058, $\mu_\alpha$~=~$-38.93$~mas~yr$^{-1}$; 
 $\delta_{2000}$~=~$-70$\degr~11\arcmin~41\farcs2363, $\mu_\delta$~=~$+0.29$~mas~yr$^{-1}$).  
The phase center of the observations was subsequently corrected using the CASA task, {\tt fixvis}.  
The continuum bandwidth amounted to 1.48 and 1.83~GHz 
averaged at 302 and 346~GHz.  
Continuum and line imaging were performed using 
the CLEAN algorithm with Briggs weighting (robust~=~0.5) resulting in synthetic beam sizes of 
$1\farcs0 \times 0\farcs48$ 
(23\textdegree) and $0\farcs95 \times 0\farcs42$ (38\textdegree) at 302 and 346~GHz.  
The synthesized beam is elongated perpendicular to the 
major axis of the disk owing, in part, to the low declination of the source (-70\textdegree).  
The continuum was subtracted from line-containing channels using the CASA task, {\tt uvcontsub}, 
in advance of imaging the CO emission.  
The achieved rms for the continuum was 0.4 and 
0.5~mJy~beam$^{-1}$ at 302 and 346~GHz, respectively, with 
an rms of 19 mJy~beam$^{-1}$~channel$^{-1}$ attained for the CO-containing channels. 
   
\section{RESULTS}
\label{results}

\begin{figure*}[!htp]
\centering
\includegraphics[width=0.618\textwidth,clip]{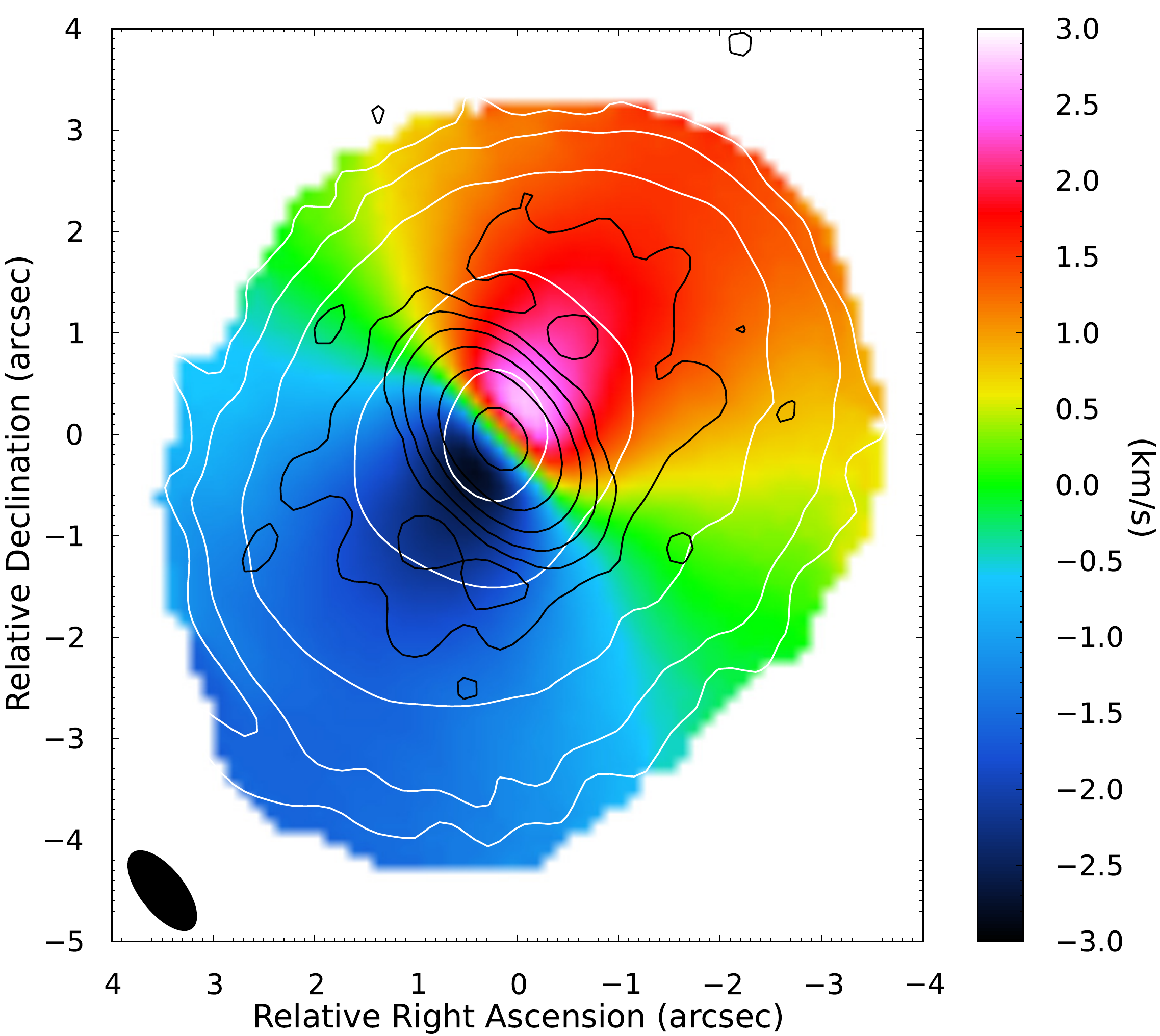}
\caption{First-moment map of CO J=3--2 emission (colour map) overlaid with  
integrated intensity contours (in white) and 870~$\mu$m continuum emission contours (in black). 
The intensity contours correspond to 3, 10, 30, 100, and 300 times the rms (30~mJy~beam$^{-1}$~km~s$^{-1}$)  
and the continuum contours correspond to 3, 10, 30, 100, 300, and 1000 times the rms (0.5 mJy~beam$^{-1}$). 
The CO integrated intensity reaches 5\% of its peak value at $\approx$~20~$\times$~rms, whereas the continuum 
emission reaches 5\% at $\approx$~60~$\times$~rms.
The synthesized beam is the same for both observations.}
\label{figure1}
\end{figure*}

Figure~\ref{figure1} presents the CO $J$=3--2 first moment map overlaid with 
contours of the integrated intensity and the continuum emission at 870~$\mu$m.    
The integrated intensity was determined between $-12$ and $+12$~km~s$^{-1}$ 
relative to the source velocity (constrained by these data to 5.7~km~s$^{-1}$), 
corresponding to channels containing significant emission ($\gtrsim$~3$\sigma$).   
The CO emission is detected with a peak signal-to-noise of 163 in the channel maps.  
The continuum emission is detected with a peak signal-to-noise of 1525 and 1320
and a total flux density of 0.980 and 1.240~Jy (summing over all flux $\gtrsim$~3~$\sigma$) 
at 302 and 346~GHz, respectively.  
The estimated absolute flux calibration uncertainties are $\approx$~10\%.    
These flux densities are consistent with previous mm observations \citep{henning94,wilner03} and yield a 
dust spectral index ($F_\nu$~$\propto$~$\nu^{\beta+2}$), $\beta$~$\approx$~0.7--0.8 between 
3.4 and 1.0~mm, that falls to $\approx$~$-0.4$ between 1.0~mm and 870~$\mu$m, indicating that 
the continuum emission is entering the optically thick regime at submillimeter wavelengths.  
The total dust mass, M$_\mathrm{dust}$~$\approx$~$D^2 F_\nu / \kappa_\nu B_\nu(T_\mathrm{dust})$, 
is $\approx$~0.035~M$_\mathrm{Jup}$, assuming
$\kappa_\nu = 10$~cm$^2$~g$^{-1}$ at 300~GHz, and $T_\mathrm{dust} = 60$~K 
\citep[see, e.g.,][]{andrews11,bruderer12}.  
  
\begin{figure}[!htp]
\centering
\includegraphics[width=0.5\textwidth,clip]{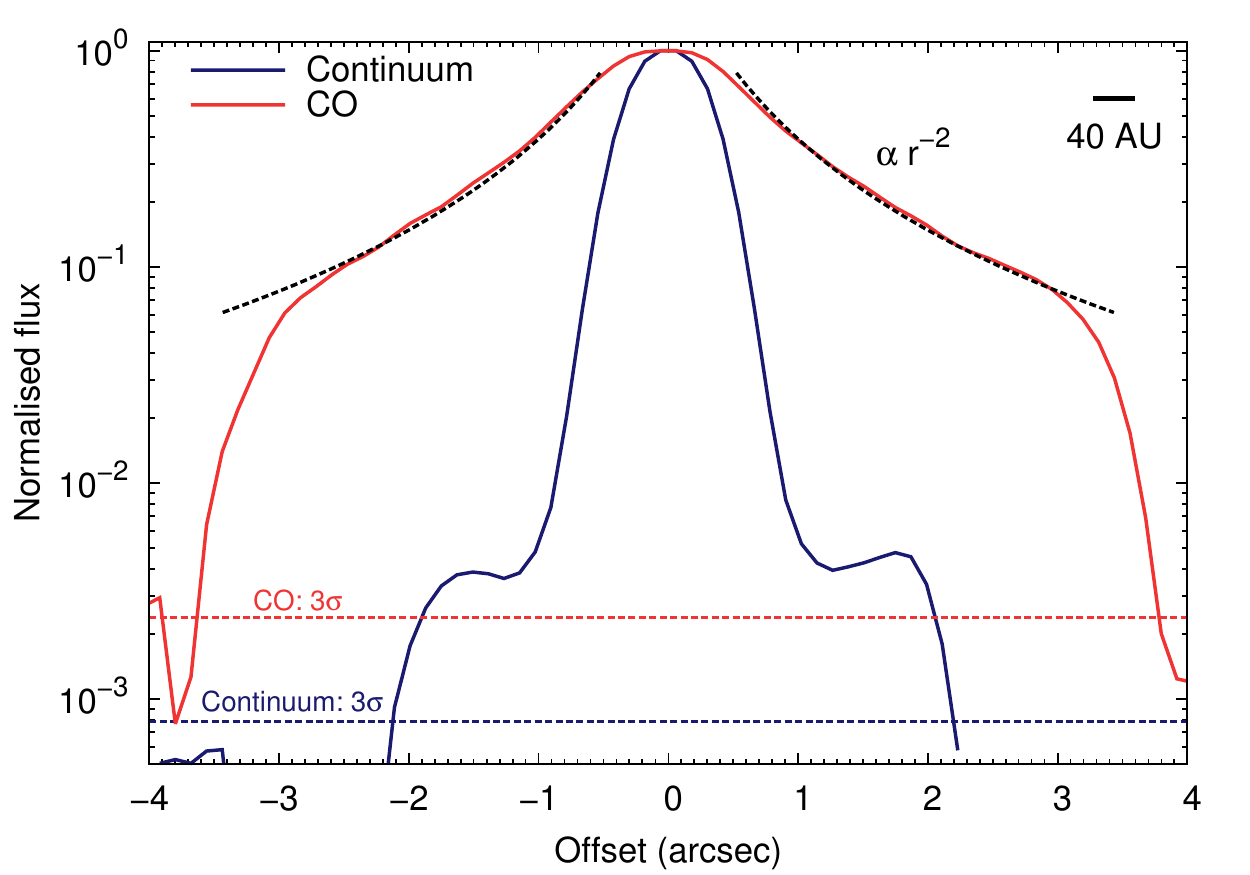}
\caption{Normalized continuum flux at 346~GHz (blue lines) and CO J=3--2 (345.795~GHz) integrated intensity (red lines) 
along the major axis of the disk, and respective 3$\sigma$ rms values (dotted lines).  
The dashed black lines show a $r^{-2}$ power law overlaid on the CO integrated intensity. 
The beam size is represented by the thick horizontal line in the top right-hand corner.}
\label{figure2}
\end{figure}

Figure~\ref{figure2} shows the continuum flux density at 346~GHz 
and CO integrated intensity along the major axis of the disk.  
The data confirm the radius of the molecular disk, 390$\pm$20~AU (the error 
corresponds to half the width of the synthesised beam).  
The CO emission is similar in extent to the scattered light images from 
\citet{ardila07} suggesting that the molecular gas and micron-sized grains are well mixed in the disk atmosphere.  
The CO brightness distribution follows a $r^{-2}$ behaviour similar
to that seen for the micron-sized grains.    
The size of the molecular disk is approaching the largest resolvable angular scale; 
hence, the drop beyond $3\arcsec$ may be caused by spatial filtering. 
However, the total integrated CO flux in these data is 151~Jy~km~s$^{-1}$ which is around 
92\% of the flux measured with APEX \citep{panic10}. 
Hence, it is unlikely that the disk extends significantly beyond the radius derived here.     
The mm continuum emission extends to only 230$\pm$20~AU and   
has two components: strong emission from the inner disk ($\lesssim 1\arcsec$) and a weaker
outer component (1\arcsec--2\farcs2) with a peak flux density $\approx$~4--5\% of the central flux.  
The self-calibration procedure (using a mask containing only the strong continuum component) 
significantly increased the dynamic range of the observations improving the peak signal-to-noise at 
346 GHz from 150 to 1320, allowing the weak extended emission to be revealed.

All subsequent analysis is conducted in the visibility domain.  
This allows a search for evidence of gaps or cavities which are not visible in the images.    
As a first step, the CASA task {\tt uvmodelfit} was used to fit the continuum 
visibilities assuming the emission arises from a elliptical disk.  
This resulted in an inclination of 44$\pm$3\textdegree and a position angle 
(measured East from North) of 146$\pm$4\textdegree, respectively, in excellent 
agreement with previous observations 
\citep[see, e.g.,][]{pantin00,augereau01,grady01,ardila07,panic10}. 

\begin{figure*}[!htp]
\subfigure{\includegraphics[width=0.5\textwidth,clip]{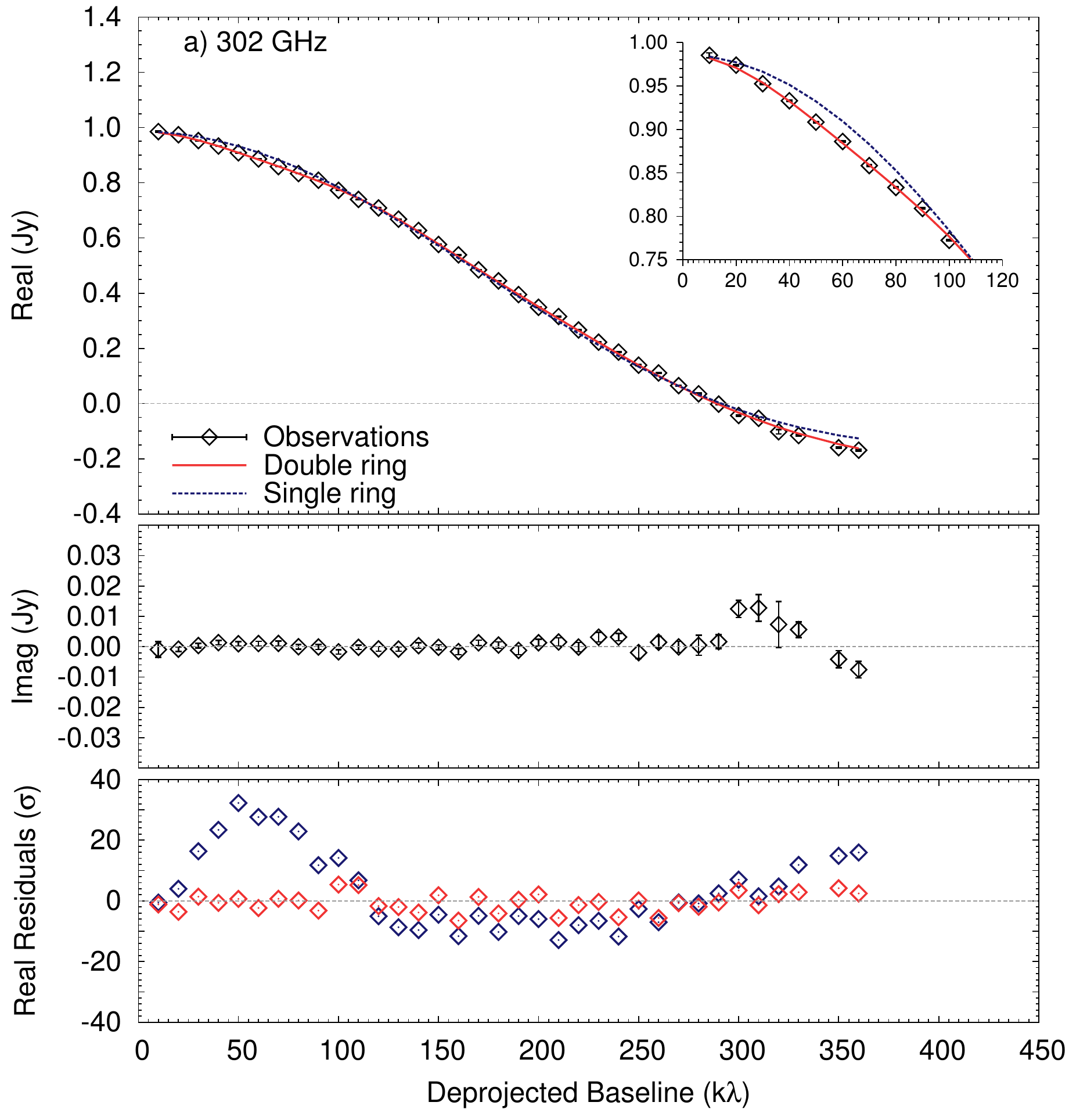}}
\subfigure{\includegraphics[width=0.5\textwidth,clip]{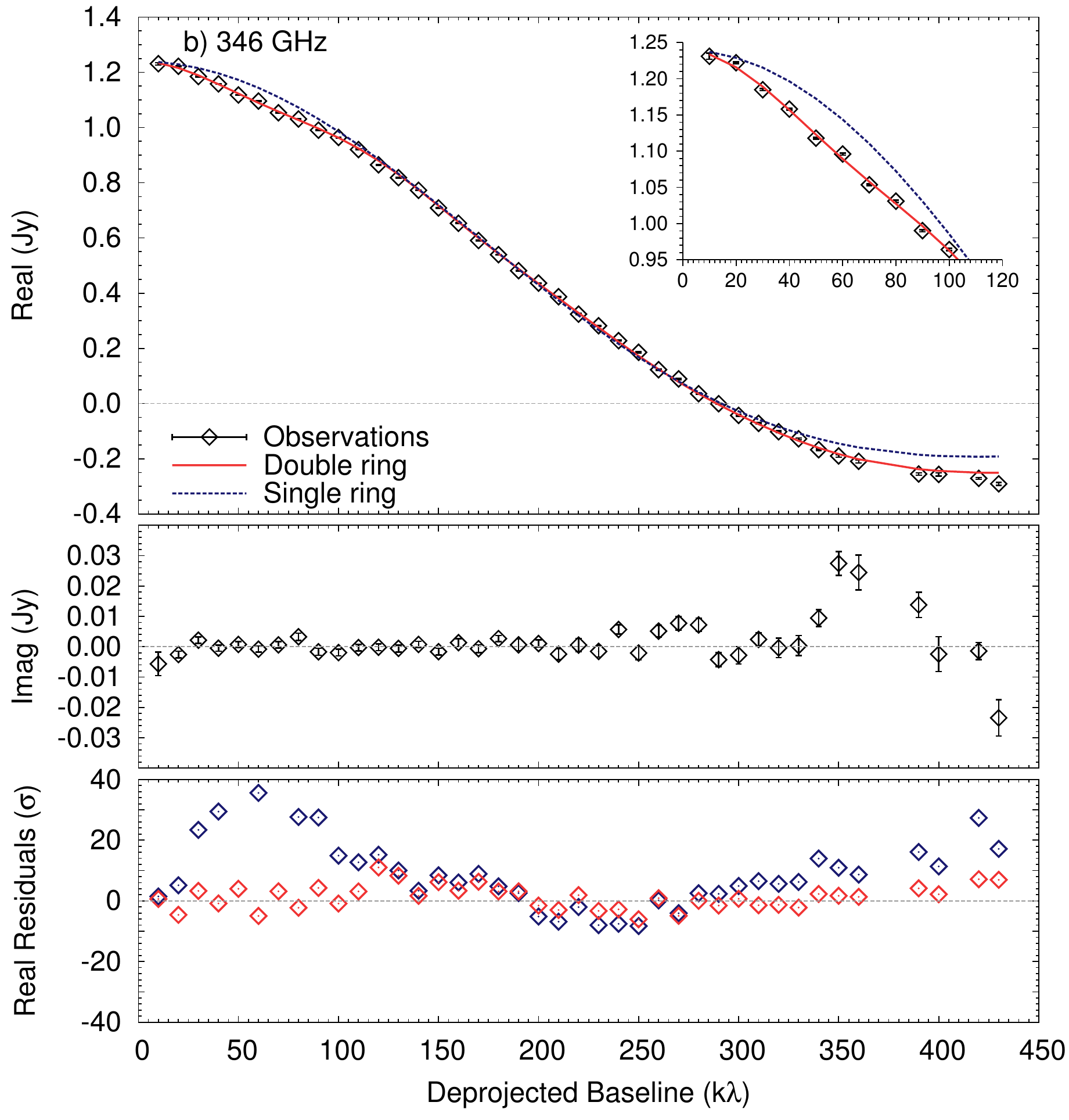}}
\caption{Visibilities as a function of the deprojected 
baseline overlaid with the best-fit ``single-ring" (blue lines) and ``double-ring" (red lines) models. 
Model residuals are shown in the bottom panel.}
\label{figure3}
\end{figure*}

Without any further knowledge on the structure we assume a circular-symmetric 
surface-brightness distribution. 
Visibilities of such distributions depend 
only on the deprojected baseline length, 
$r_{uv} = \sqrt{u_\phi^2\cos^2 i + v_\phi^2}$, where 
$u_\phi = u\cos\phi  + v\sin\phi$ and 
$v_\phi =  -u\sin\phi + v\cos\phi$ assuming the $u$-axis is aligned with right ascension \citep[see, e.g.,][]{berger07}.  
Here, $(u,v)$ are the observed visibility coordinates, 
$i$ is the source inclination, and $\phi$ is the disk position angle.   
Figure~\ref{figure3} presents the binned visibilities 
(in 10~k$\lambda$ bins) as a function of $r_{uv}$. 
The error bars correspond to the standard error in each bin.
The imaginary components show very small scatter around zero $\lesssim$~250--300~k$\lambda$, 
confirming the assumption of a symmetric brightness distribution {\em a posteriori} 
(a point-symmetric brightness distribution has zero imaginary components).
For $r_{uv}\gtrsim 250\mathrm{-}300$~k$\lambda$, 
the scatter in the imaginary components increases, which may indicate 
an asymmetry in the continuum emission; however, this may also be caused by coarser the $uv$ 
coverage at long baselines. 
Higher spatial resolution observations are needed to confirm 
any asymmetry at small spatial scales. 
The real components of the visibilities decrease as a function 
of the deprojected baseline indicating the continuum emission is resolved 
and there is a zero crossing (null) at 290~k$\lambda$.  
The Fourier transform of an infinitesimally narrow ring is 
a Bessel function of the first kind, $J_0$: a null suggests the emission 
originates from a ring with a finite width \citep[e.g., ][]{berger07,hughes07}.  
\citet{pineda14} determine a null position at 250~k$\lambda$.    
This is likely due to an incorrect deprojection related to the convention of the 
direction of the $u$-axis relative to right ascension \citep{berger07,hughes07}. 

For a ring, the real component of the visibilities is given by
\begin{equation}
V_{\mathrm{Re}}(r_{uv}) = 2\pi \int_{0}^{\infty} I(\theta)J_0(2 \pi r_{uv}\theta )\theta \, \mathrm{d}\theta
\end{equation}
\citep{berger07}.
The intensity, $I(\theta)$, is modelled as a power-law,
\begin{equation}
I(\theta) = 
\begin{cases} 
C \cdot \theta^{-\gamma} \quad \mathrm{for} \quad \theta \ge \theta_\mathrm{in} \quad \mathrm{and} \quad \theta \le \theta_\mathrm{out} \\
0 \quad \quad \mathrm{otherwise}.
\end{cases}
\end{equation}
The flux scaling factor, $C$, is determined using the total observed flux, 
$V_{\mathrm{Re}}(0)$, i.e, $C = V_\mathrm{Re}(0)/\int_{0}^{\infty} I(\theta)J_0(0)\theta \mathrm{d}\theta$.  
$\theta_\mathrm{in}$ and $\theta_\mathrm{out}$ 
were varied between 0 and 50~AU and 20 and 400~AU, respectively, 
for $\gamma$~=~0, 1, and 2, using a small step size (1~AU) to adequately 
sample the parameter space.

\begin{figure*}[!htp]
\subfigure{\includegraphics[width=0.5\textwidth,clip]{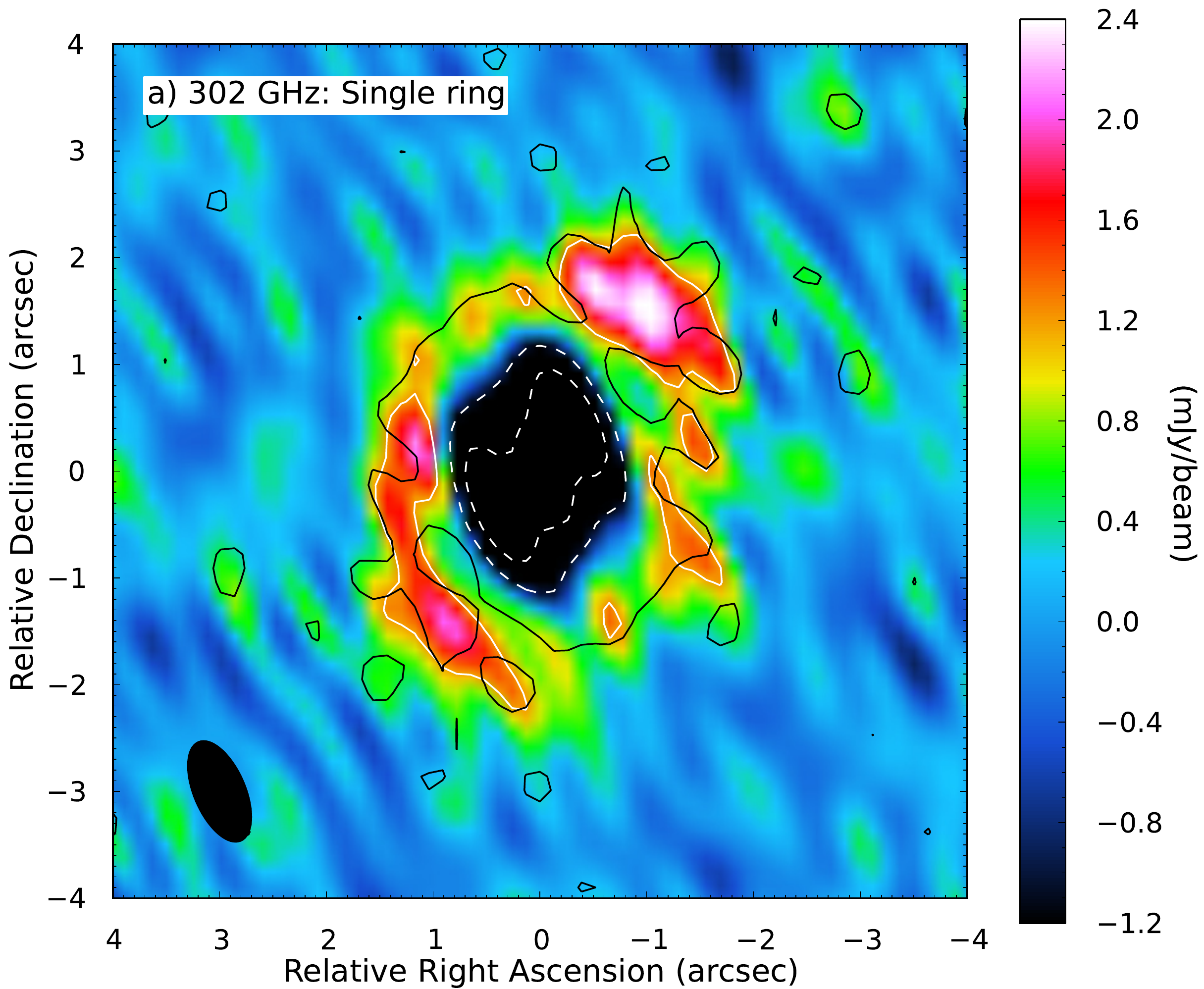}}
\subfigure{\includegraphics[width=0.5\textwidth,clip]{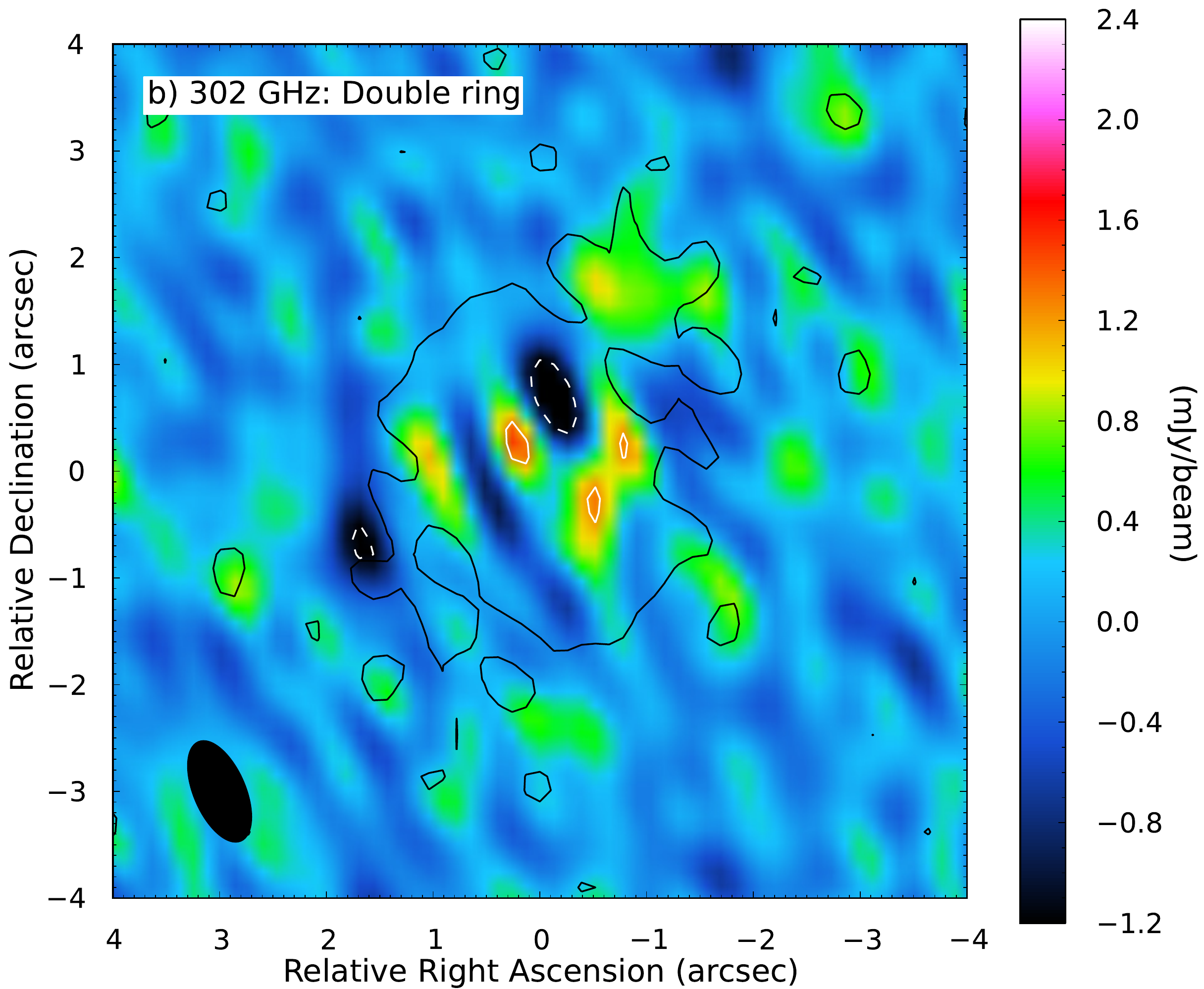}}
\subfigure{\includegraphics[width=0.5\textwidth,clip]{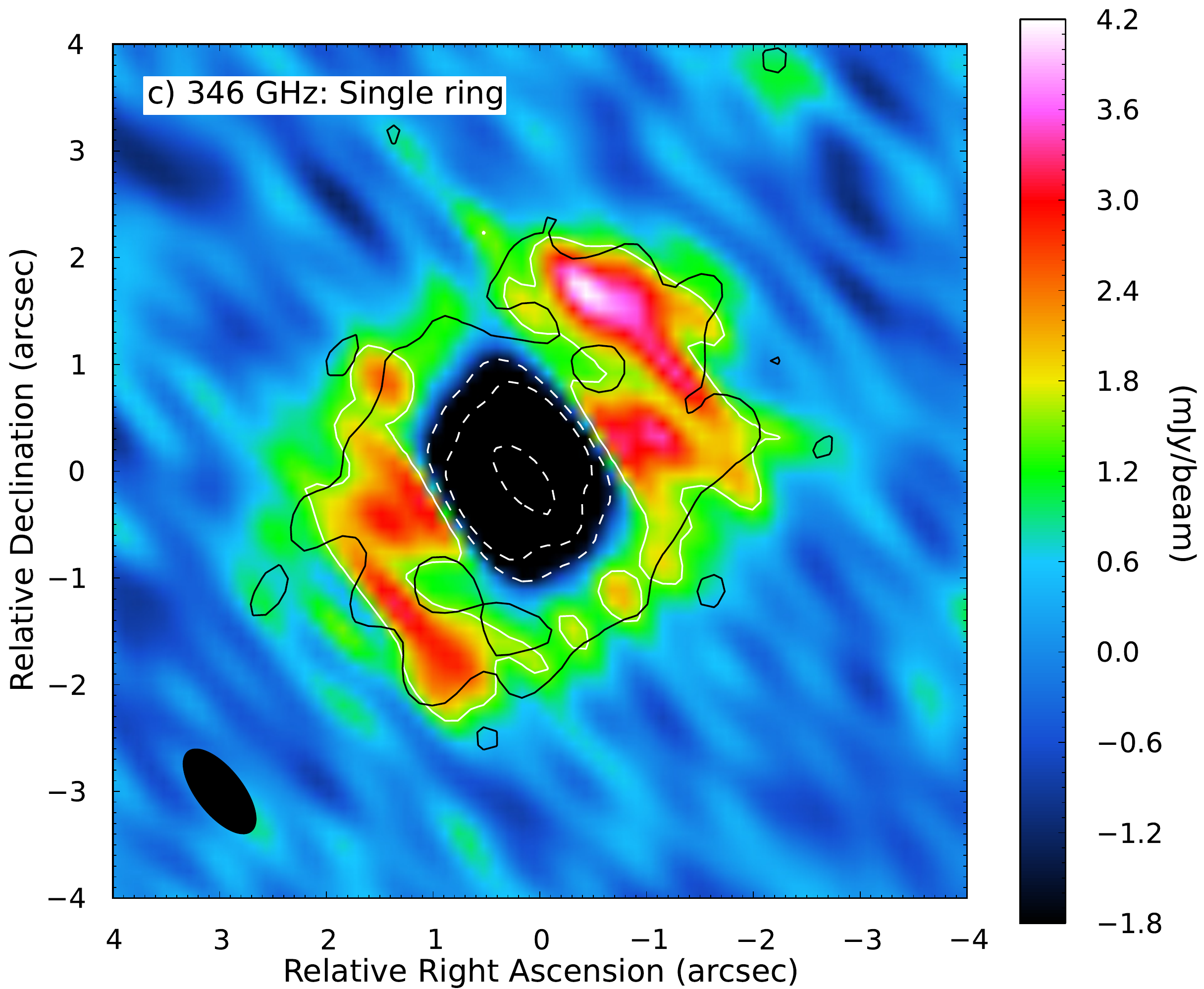}}
\subfigure{\includegraphics[width=0.5\textwidth,clip]{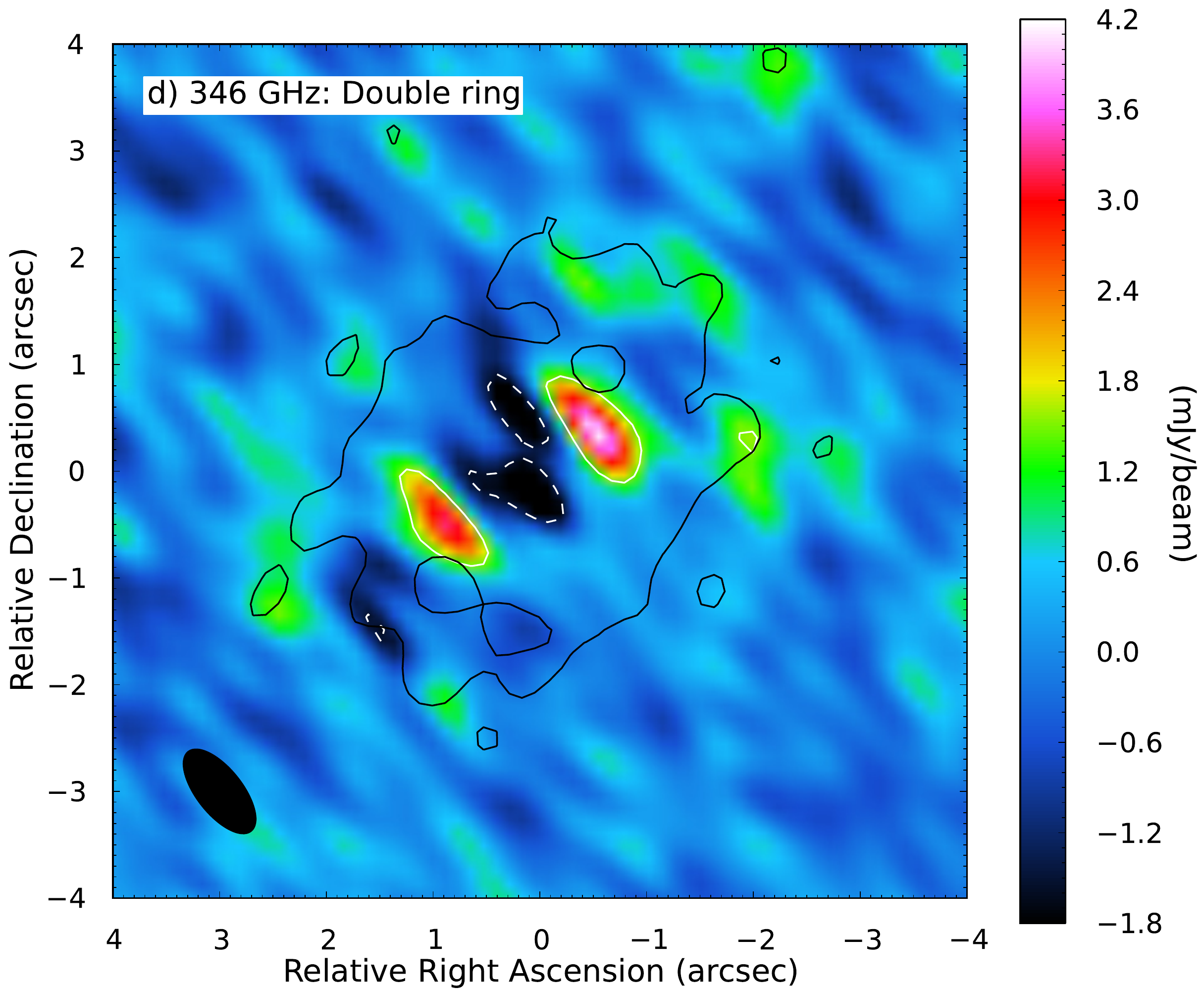}}
\caption{Residual images (colour map) and contours 
(white lines) overlaid with observed 3$\sigma$ contour (black lines) for the 
single-ring (left-hand panels) and double-ring (right-hand panels) models. 
The colorbar scale is truncated to highlight the significance of the residual outer ring.  
The dashed contours indicate negative residuals (-3, -10, and -30$\sigma$).}
\label{figure4}
\end{figure*}

The best-fit model has an inner and outer radius of 16 and 51~AU, 
and a power-law index of 2 (see Figure~\ref{figure3}).  
This model corresponds to a deep global minimum in the $\chi^2$ value indicating that the 
estimated uncertainties are smaller than the step size of the grid (1~AU). 
The model residuals were imaged using an identical $uv$ coverage as 
the observations (see Figure~\ref{figure4}).  
The residuals in both the visibility and image domains 
are large ($\gg$~3$\sigma$) indicating a poor fit. 
The images reveal significant extended, weak 
continuum emission (peak residuals~=~6~--~8$\sigma$) . 
\citet{pineda14} do not see this extended emission because 
no self calibration of the data was performed. 

To include this more extended component, the model was adapted to include i)   
a compact ring with a gaussian brightness distribution, 
\begin{equation}
I(\theta) = C \exp{\left( \frac{-(\theta - \theta_\mathrm{peak})^2}{2\theta_\mathrm{width}^2}\right)},
\end{equation}
\citep{perez14} 
and ii) an extended disk/ring with a flat brightness distribution ($\gamma$~=~0).  
A low-resolution grid was run (5~AU) to determine 
the location of the global minimum in the $\chi^2$ value.  
A subsequently denser grid was run in which
$\theta_\mathrm{peak}$ and $\theta_\mathrm{width}$ were varied 
between 20 and 50~AU and 5 and 20~AU, respectively, and   
$\theta_\mathrm{in}$ and $\theta_\mathrm{out}$ between 10 and 200~AU   
and 100 and 400~AU.  
This grid included models composed of both overlapping and distinct rings.
A small step size of 0.5~AU was chosen to allow quantification of the 
errors via Bayesian inference.  
For simplicity, the total flux contribution from the compact and extended  
components were fixed at 0.962 and 0.024~Jy at 302~GHz and 
1.190 and 0.048~Jy at 346~GHz, respectively.  
This was set by the flux in the residual images. 

The visibilities are best reproduced by a  
compact ring with a peak brightness at 26~AU and a FWHM of 21~AU and  
an outer ring with a width of 75$\pm$3~AU centered at 190$\pm$3~AU 
(see Figure~\ref{figure3}).   
The data exclude overlapping rings in favour of two distinct rings of emission.
The estimated dust masses for the inner and outer rings  
are $\approx$~2.5~$\times$~10$^{-2}$ and 
$\approx$~1.4~$\times$~10$^{-3}$~M$_\mathrm{Jup}$,  
assuming disk temperatures of 80 and 40~K at 
$\approx$~30 and $\approx$~190~AU \citep[see, e.g.,][]{bruderer12}.

Figure~\ref{figure4} shows the residual images for the ``double-ring" model.   
The peak residuals at 302 and 346~GHz are 1.5~mJy (3.8$\sigma$)  
and 4.1~mJy (8.2$\sigma$).    
However, these are restricted to small regions and 
are likely owing to deviations from circular symmetry also suggested by non-zero 
imaginary components on long baselines (see~Figure~\ref{figure3}).   

\section{DISCUSSION}
\label{discussion}

Previous observations show that the mm-sized grains are not 
necessarily cospatial with the molecular gas in protoplanetary disks 
\citep[][]{isella07,andrews12,degregoriomonsalvo13}.    
This can be explained by {\em radial drift}:   
dust grains feel a drag force as they move through 
the sub-Keplerian gas causing a loss of angular momentum and  
migration inwards towards the star \citep[e.g.,][]{birnstiel10}.  
When a massive companion opens a gap in the disk \citep[e.g.,][]{kley12}, this halts 
the migration of grains because of the presence of a positive pressure 
gradient at the outer edge of the gap.  
Grains can accumulate and grow in this ``pressure trap" with the peak 
and structure of the pressure profile dependent 
on the disk viscosity and the location and mass of the companion \citep{pinilla12}.  

Observations of HD~100546 support the presence of a 
close-in companion \citep[e.g.,][]{acke06,liskowsky12,brittain13}.  
\citet{mulders13} derived a lower limit of 20~M$_\mathrm{Jup}$ for the 
companion mass and constrained the disk viscosity to 
$\alpha_\mathrm{turb} \gtrsim 2\times 10^{-3}$.
A potential protoplanet has also been observed at 68$\pm$10~AU \citep{quanz13}. 
The ALMA observations suggest the mm-sized grains are located in two rings:  
one between the proposed companions and the other beyond the outer protoplanet.  
To investigate the influence of companions on the dust evolution in HD~100546, 
we model the dust growth and migration for two scenarios \citep{birnstiel10,pinilla12}: 
(a) a 20~M$_\mathrm{Jup}$ companion at 10~AU only, and 
(b) both a 20~M$_\mathrm{Jup}$ companion and a protoplanet (15~M$_\mathrm{Jup}$) at 68~AU.  
We assume an initial particle size, 1$\mu$m, 
a disk viscosity, $\alpha_\mathrm{turb} = 2\times 10^{-3}$, 
a stellar mass, 2.4~$\mathrm{M}_\odot$, and a dust mass, $5.0 \times 10^{-4}$~$\mathrm{M}_\odot$ \citep{mulders13}.  
The model from \citet{mulders13} is extrapolated to larger radii (400~AU) 
using a power-law and assuming a gas-to-dust mass ratio of 100.  
The younger protoplanet is injected into the simulations at 1~Myr. 

\begin{figure*}[!htp]
\includegraphics[width=\textwidth]{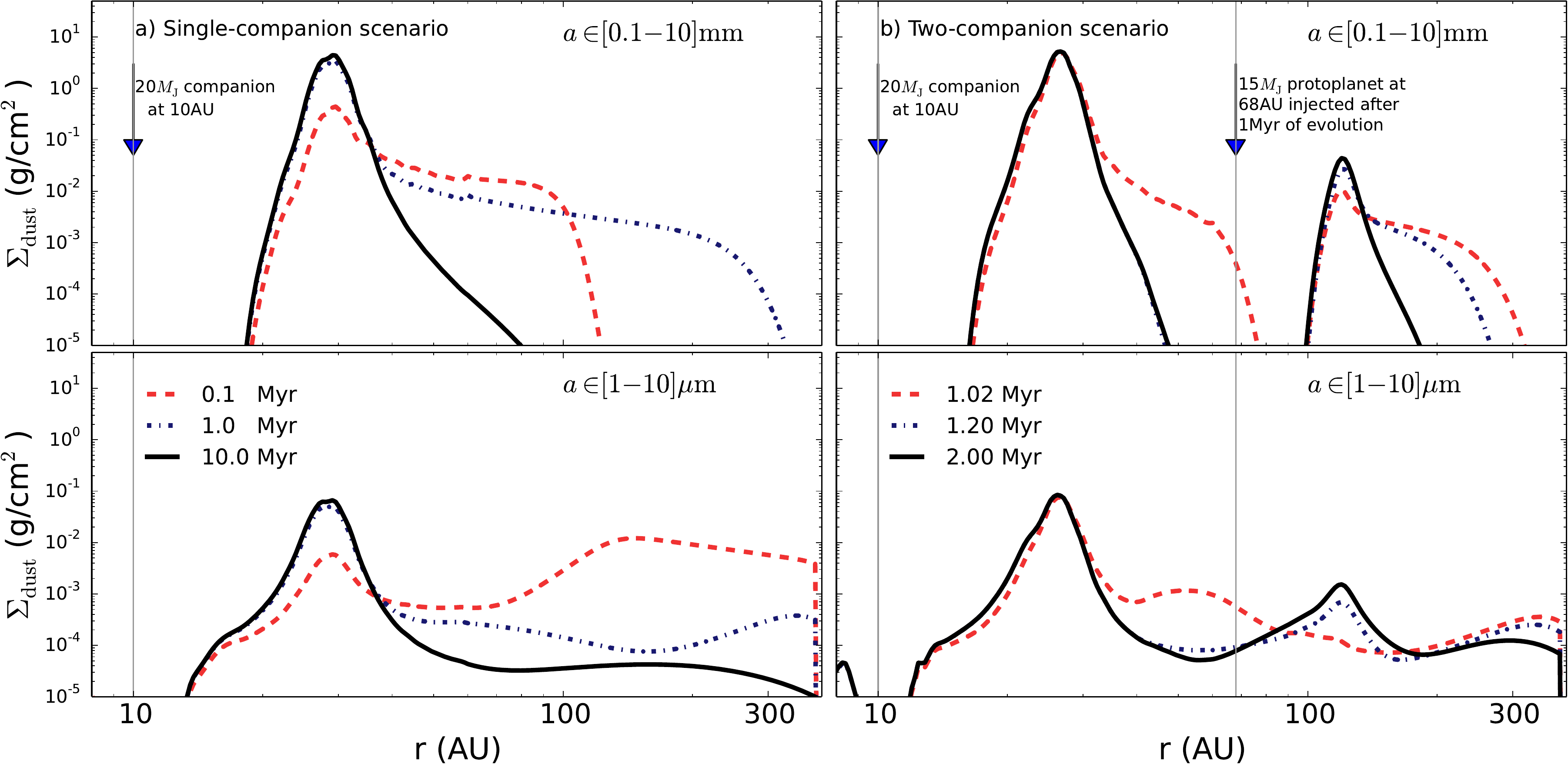}
\caption{Surface density of mm-sized and $\mu$m-sized grains (top and bottom rows, respectively)
for (a) the single-companion scenario, and (b) the two-companion scenario (left- and right-hand panels, respectively).  
The surface density for case (a) is plotted after 0.1, 1.0, and 10~Myr (red, blue, and black lines, respectively) of evolution. 
The surface density for case (b) is plotted after 0.02, 0.2, and 1~Myr (red, blue, and black lines, respectively) 
of evolution following the injection of the protoplanet at 1~Myr.}
\label{figure5}
\end{figure*}

Figure~\ref{figure5} presents the surface density of mm-sized and $\mu$m-sized
grains at different simulation times.  
For the single-companion scenario, long evolution times are required, $\approx$~10~Myr, 
for the grains to grow to mm sizes in the outer regions (100--400~AU) 
and migrate inwards to accumulate in a radial pressure trap with a peak at 
$\approx$~30~AU and a width of $\approx$~20~AU.  
For the two-companion scenario, the surface density 
decreases sharply in the region between the two companions upon introduction 
of the protoplanet after 1~Myr of dust evolution.
The resulting steep pressure gradient causes 
grains to migrate inwards on shorter timescales, $\approx$~0.2~Myr.    
The injection of the protoplanet at 68~AU generates a second ring 
at $\gtrsim$~100~AU with a surface density $\approx$~100--1000 times 
lower than that for the inner ring.   
Around 1.0~Myr of additional evolution is required 
for this ring to narrow to a width $\lesssim$~100~AU.  
These results are qualitatively in agreement with the ALMA observations 
which also show a contrast of $\sim$~100 between emission from the inner and outer ring.  
The $\mu$m-sized grains extend from $\approx$~12--13~AU to $\approx$~400~AU   
which is consistent with scattered light observations and SED models of the source.  

These observations and simulations support the presence of a massive companion 
orbiting within the inner gap and a protoplanet embedded within the outer disk.  
Numerical models of dust evolution including two companions recreate the inner 
ring of mm emission and the extended weaker emission seen in the ALMA data.  
Particle trapping by the inner companion alone cannot explain the nature 
of the outer ring. 

\acknowledgments

This paper makes use of the following ALMA data: ADS/JAO.ALMA\#2011.0.00863.S. 
ALMA is a partnership of ESO (representing its member states), 
NSF (USA) and NINS (Japan), together with NRC (Canada) and NSC and 
ASIAA (Taiwan), in cooperation with the Republic of Chile. 
The Joint ALMA Observatory is operated by ESO, AUI/NRAO and NAOJ.  
The authors thank E.~F.~van~Dishoeck, C.~P.~Dullemond, N. van~der~Marel, 
and M.~Schmalzl for useful discussions, and G.~D.~Mulders for sharing the results of his 
hydrodynamical simulations.  
C.~W. acknowledges support from the 
Netherlands Organisation for Scientific Research (NWO, program number 639.041.335).  
This work was also supported by EU A-ERC grant 291141 CHEMPLAN and a KNAW prize.  
T.~B. acknowledges support from NASA Origins of Solar Systems grant NNX12AJ04G.
Astrophysics at QUB is supported by a grant from the STFC.  
M.~R.~H., A.~J., and G.~S.~M. acknowledge support from the Netherlands Organization for 
Scientific Research (NWO) to Allegro, the European ALMA Regional Center node in the Netherlands.  


\end{document}